\documentclass{article}
\usepackage{epsfig}

\usepackage{psfrag}

\psfrag{p}{$p=(1+\xi)P-\Delta_\perp/2$}
\psfrag{pi}{$\pi$}
\psfrag{1-xi}{$1-\xi$}
\psfrag{1+xi}{$1+\xi$}
\psfrag{b}{${\bf b}$}
\psfrag{xib/(1+xi)}{$ \frac{\xi}{1+\xi} {\bf b}$}
\psfrag{xi/(1-xi)b}{$\frac{\xi}{1-\xi} {\bf b}$}
\psfrag{u}{$u$}
\psfrag{d}{$d$}
\psfrag{1/Q}{$\frac{1}{Q}$}
\psfrag{b}{${\bf b}$}
  
\psfrag{pp}{$p'=(1-\xi)P+\Delta_\perp/2$}
\psfrag{q}{$q$}
\psfrag{qp}{$q'$}
\psfrag{pip}{$\pi^+$}
\psfrag{pim}{$\pi^-$}
\psfrag{pr}{$P$}
\psfrag{apr}{$\bar P$}
\psfrag{g}{$\gamma$}
\psfrag{gs}{$\gamma^*$}
\psfrag{u}{$u$}
\psfrag{db}{$\bar d$}
\psfrag{ep}{$e^+$}
\psfrag{em}{$e^-$}
\psfrag{d}{$d$} 
\psfrag{a}{$a$}
\psfrag{TH}{$T_H$}
\psfrag{DA}{$DA$}
\psfrag{TDA}{$TDA$}
\psfrag{ga}{$\gamma$}
\psfrag{x+xi}{$x+\xi$}
\psfrag{x-xi}{$x-\xi$}

\usepackage{amssymb} 
\usepackage{amsmath} 
\usepackage{amsfonts} 
\usepackage{cite} 

\newcommand{\beq}[1]{
\begin{equation}\label{#1}}
\newcommand{\eeq}{\end{equation}}
\newcommand{\bea}[1]{
\begin{eqnarray}\label{#1}}
\newcommand{\eea}{\end{eqnarray}}
\begin{document}

\begin{titlepage}

\begin{center}

{\LARGE \bf
The QCD analysis of hadron-antihadron $\to \gamma^* \gamma$ in 
the forward region and 
related processes\footnote{
Presented at PHOTON2005 International Conference on 
the Structure and Interactions of the Photon, Warsaw 31.08-04.09.2005
 by L. Szymanowski.}}

\vspace{1cm}

{\sc B.~Pire}${}^{1}$,
{\sc L.~Szymanowski}${}^{3,4}$ 
\\[0.5cm]
\vspace*{0.1cm} ${}^1${\it
CPhT, {\'E}cole Polytechnique, F-91128 Palaiseau, France\footnote{
  Unit{\'e} mixte C7644 du CNRS.} \\[0.2cm]
\vspace*{0.1cm} ${}^2$  {\it
 So{\l}tan Institute for Nuclear Studies,
Ho\.za 69,\\ 00-681 Warsaw, Poland
                       } \\[0.2cm]
\vspace*{0.1cm} ${}^3$ {\it
Universit\'e  de Li\`ege,  B4000  Li\`ege, Belgium  } \\[1.0cm]
}

\end{center}
\vskip2cm

A QCD analysis of the reactions 
$A B \to \gamma^* \gamma$ and $\gamma^* \gamma \to A B  $
in the forward region shows that they can be factorized 
in a way which is quite similar to the 
frameword developped for deeply virtual Compton scattering. 
The generalized parton 
distributions (GPDs) related to this latter process 
being replaced by new non perturbative
hadronic matrix elements, the transition distribution amplitudes (TDAs).
\vskip1cm

\vspace*{1cm}

\end{titlepage}

\section{Introduction}
The recent successes of the colinear factorization approach in exclusive hard processes 
motivates further studies where a scaling  regime may be reached for amplitudes
which factorize at leading twist  into 
a short-distance matrix element  and  long-distance dominated  matrix elements between
hadron (or hadron and vacuum) states. We thus 
propose \cite{PS2} to write the $AB \to \gamma^*\gamma$ amplitude  
in the scaling regime with large $Q^2$ and $s$ with fixed $Q^2/s$ and 
$-t<<Q^2$ as 
\begin{equation}
{\cal M} (Q^2,  \xi, t)= \int dx dy \phi_{A}(y_i,Q^2)
T_{H}(x_i, y_{i}, Q^2) T(x_{i}, \xi, t, Q^2)\;,
\label{amp}
\end{equation}
where $\phi_{A}(y_i,Q^2)$ is the A-hadron distribution amplitude,
$T_{H}$ the hard scattering amplitude, calculated in the colinear approximation and $T(x_{i}, \xi, t, Q^2)$
the new $B \to \gamma$ TDAs. 
$x_i$ and $y_i$ are light cone momentum
fractions of quarks in hadrons.
The same formula applies as well to the time reversed reaction
$\gamma^* \gamma \to A B  $. 
Backward VCS can also be
 described 
in the same framework.

\section{$\gamma \to \pi $ Transition Distribution Amplitude}
The $ \gamma  \to\pi $ TDAs are defined as  \cite{PS2}
\noindent
\begin{eqnarray}
  \label{Vpi} 
&&\hspace*{-1cm} \int \frac{d z^-}{2\pi}\, e^{ix P^+ z^-}
\langle     \pi^-(p')|\, \bar{d}(-z/2)[-z/2,z/2]\,\gamma^\mu
\,{u}(z/2)
\,|\gamma(p,\varepsilon) \rangle \Big|_{z^+=  z_T=0} \nonumber\\
&&=\frac{1}{P^+} \frac{i\;e}{f_\pi}\epsilon^{\mu \nu \rho
\sigma}\varepsilon_{\perp\nu}
P_\rho \Delta_{\perp \sigma}\;V(x,\xi,t)\;, \nonumber
  \end{eqnarray}
\noindent
\begin{eqnarray}
  \label{Api} 
&&\hspace*{-1cm}\int \frac{d z^-}{2\pi}\, e^{ix P^+z^-}       
\langle       \pi^-(p')|\,   
 \bar{d}(-z/2)[-z/2,z/2]\,\gamma^\mu 
\gamma_{5}\,{u}(z/2)    \,|\gamma(p,\varepsilon)    \rangle 
\Big|_{z^+=  z_T=0}
\nonumber \\ 
&&
  = \frac{1}{P^+} \frac{e}{f_\pi} (\vec 
\varepsilon \cdot \vec \Delta) P^\mu \;A(x,\xi,t)\;, \nonumber
 \end{eqnarray}
\begin{eqnarray}
  \label{T} 
&&\hspace*{-1cm}\int \frac{d z^-}{2\pi}\, e^{ix P^+z^-}
\langle       \pi^-(p')|\,  \bar{d}(-z/2)[-z/2,z/2]\,
\sigma^{\mu \nu} \,{u}(z/2)  \,|\gamma(p,\varepsilon)    
\rangle \Big|_{z^+=  z_T=0}  \nonumber\\
&& = \frac{e}{P^+} \epsilon^{\mu \nu \rho  
\sigma} P_\sigma \left[ 
 \varepsilon_{\perp \rho} 
 T_1(x,\xi,t) -   \frac{1}{f_\pi} (\vec \varepsilon \cdot \vec 
\Delta)   \Delta_{\perp \rho} T_2(x,\xi,t) \right]  \;,
  \end{eqnarray}
where the first two TDAs, $V(x,\xi,t)$ and $A(x,\xi,t)$ are chiral even 
and the latter ones, $T_i(x,\xi,t), i=1,2$, are chiral 
odd. In definitions (\ref{Vpi})
we include the 
Wilson line 
$[y;z]  \equiv   {\rm  P\   exp\,}  \left[ig(y-z)\int_0^1\!dt\,
\,n_\mu A^\mu  (ty+(1-t)z)\right]$, which provides  the QCD-gauge
invariance for  non local operators  and equals unity in  a light-like
(axial) gauge. We also introduce standard vectors $P=(p+p')/2$ 
and $\Delta=p'-p$. $f_\pi$ is the pion decay constant. The four leading twist
TDAs are linear combinations of the four independent helicity amplitudes 
for the process $q \gamma \to q \pi^-$. 
The $x$ and $\xi $ variables, see Fig.~1, have the same meaning as in the GPDs.
The TDAs have polynomiality properties as GPDs.

\begin{figure}[h]
\begin{center}
{\includegraphics[width=7cm]{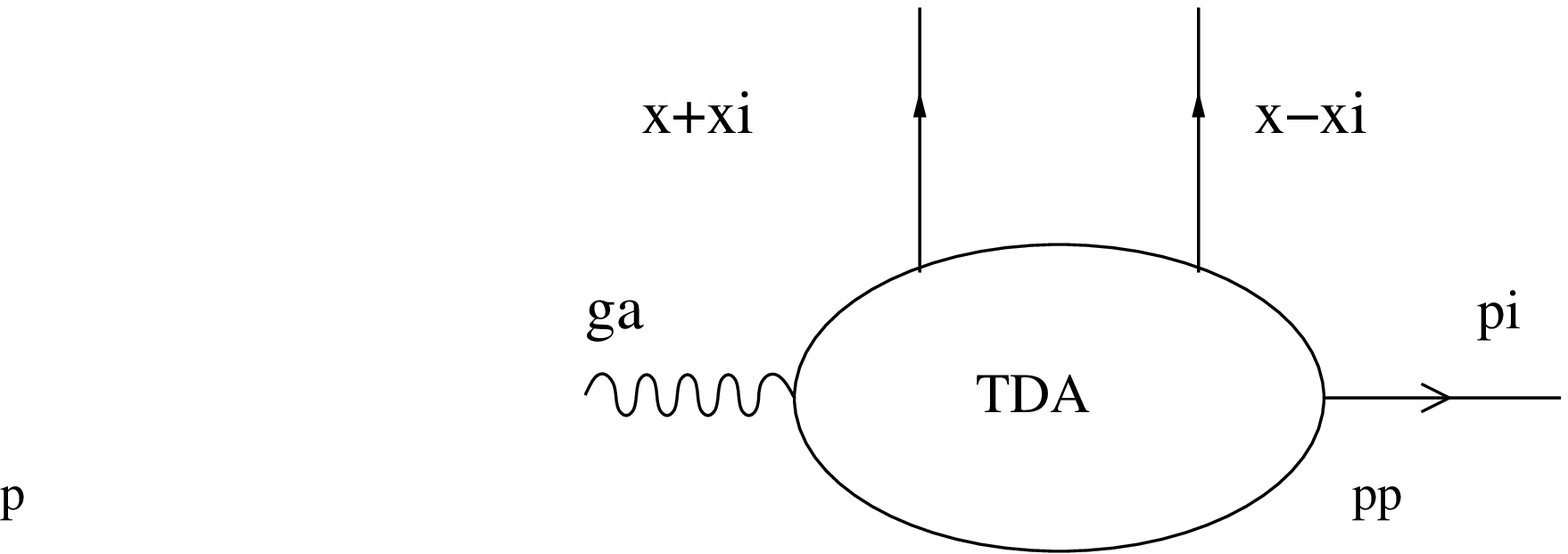}}
\caption{The  $\gamma \to \pi$ transition distribution amplitude (TDA).
}
\label{angular}
\end{center}
\end{figure}

The TDAs obey QCD evolution equations which are the GPD mixed DGLAP-ERBL 
evolution equations.
Contrarily to the case of the forward photon structure functions, we do
not expect that the pointlike nature of the photon induces new terms
in the DGLAP evolution of the $\gamma \to \pi$ TDA with respect to the
$\pi \to \pi$ GPD.

\section{Other cases}
\subsection{mesonic channels}
The $ \gamma  \to\rho $ TDAs are of direct phenomenological interest, since the reaction 
 $ \gamma^* \gamma \to\rho \rho $ has turned out to be accessible by LEP experiments \cite{LEP}, although in a 
 slightly different kinematical domain \cite{GDA}.
  Their general decomposition at leading twist can be written as
\begin{eqnarray}
&&
  \int \frac{d \kappa}{2 \pi}\,
  e^{i x \kappa P.n}
  \langle \rho^-(p', \epsilon') |\,
  \bar{d}(-\kappa  n)\, \gamma.n\, u(\kappa n)
  \,| \gamma(p, \epsilon) \rangle   
\nonumber \\
&&\hspace{-1cm} = 
H_1^{(\rho \gamma)}(x,\xi,t)\left[- (\epsilon'^*\cdot \epsilon)+\frac{8 (\epsilon'^*\cdot P)(\epsilon\cdot P)}{M^2-t}
   \right]
\nonumber \\
&&\hspace{-1cm} +H_2^{(\rho \gamma)}(x,\xi,t)\left[ (\epsilon'^*\cdot P)\frac{\epsilon\cdot n}{P\cdot n} + 
(\epsilon\cdot P)\frac{\epsilon'^*\cdot n}{P\cdot n} - \frac{(\epsilon\cdot P) (\epsilon'^*\cdot P)}{6M^2(1+\xi)}\left(-1 + \frac{24M^2(1+\xi)^2}{M^2-t}\right)
\right.
\nonumber \\
&&\hspace{-1cm} \left. - \frac{M^2-t}{48M^2(1+\xi )}\left( (\epsilon'^*\cdot \epsilon) + 
\frac{12M^2(\epsilon\cdot n)(\epsilon'^*\cdot n)}{(P\cdot n)^2}   \right)   \right]
\nonumber \\
&&\hspace{-1cm} +H_4^{(\rho \gamma)}(x,\xi,t)\left[ (\epsilon'^*\cdot P)\frac{\epsilon\cdot n}{P\cdot n} - 
(\epsilon\cdot P)\frac{\epsilon'^*\cdot n}{P\cdot n} 
- \frac{(\epsilon\cdot P) (\epsilon'^*\cdot P)}{6M^2(1+\xi)}\left(1 + \frac{24M^2(1+\xi)^2}{M^2-t}\right)
\right.
\nonumber \\
&&\hspace{-1cm} \left. + \frac{M^2-t}{48M^2(1+\xi )}\left( (\epsilon'^*\cdot \epsilon) + 
\frac{12M^2(\epsilon\cdot n)(\epsilon'^*\cdot n)}{(P\cdot n)^2}   \right)   \right]
\end{eqnarray}

\begin{eqnarray}
&&\hspace{-1cm}
  \int \frac{d \kappa}{2 \pi}\,
  e^{i x \kappa P.n}
  \langle \rho^-(p', \epsilon') |\,
  \bar{d}(-\kappa n)\, \gamma.n \gamma_5\, u(\kappa n)
  \,| \gamma(p, \epsilon) \rangle
\nonumber \\
&&\hspace{-1cm} = i \tilde H_1^{(\rho \gamma)}(x,\xi,t)\left[ -\frac{\epsilon_{\mu \alpha \beta \gamma}n^\mu \epsilon'^{*\alpha} 
\epsilon^\beta  P^\gamma}{(P\cdot n)} -\frac{2\epsilon_{\mu \alpha \beta \gamma}n^\mu \Delta^\alpha 
P^\beta (\epsilon^\gamma (\epsilon'^*\cdot P)+ \epsilon'^{*\gamma}(\epsilon\cdot P)) }{(M^2-t)(P\cdot n)}
 \right]
\nonumber \\
&&\hspace{-1cm} +i \tilde H_2^{(\rho \gamma)}(x,\xi,t)\left[ -\frac{\epsilon_{\mu \alpha \beta \gamma}n^\mu \epsilon'^{*\alpha} 
\epsilon^\beta  P^\gamma}{(P\cdot n)} +\frac{2\epsilon_{\mu \alpha \beta \gamma}n^\mu \Delta^\alpha 
P^\beta (\epsilon^\gamma (\epsilon'^*\cdot P)+ \epsilon'^{*\gamma}(\epsilon\cdot P)) }{(M^2-t)(P\cdot n)}
\right.
\nonumber \\
&& \hspace{-1cm}\left. -  \frac{\epsilon_{\mu \alpha \beta \gamma}n^\mu \Delta^\alpha 
P^\beta (\epsilon^\gamma (\epsilon'^*\cdot n)+ \epsilon'^{*\gamma}(\epsilon\cdot n)) }{(1+\xi)(P\cdot n)^2}
\right]
\nonumber \\
&&\hspace{-1cm} + i \tilde H_3^{(\rho \gamma)}(x,\xi,t)\left[ \frac{\epsilon_{\mu \alpha \beta \gamma}n^\mu \Delta^\alpha 
P^\beta (\epsilon^\gamma (\epsilon'^*\cdot P)- \epsilon'^{*\gamma}(\epsilon\cdot P)) }{M^2(P\cdot n)}
\right.
\nonumber \\
&& \hspace{-1cm}\left. +  \frac{\epsilon_{\mu \alpha \beta \gamma}n^\mu \Delta^\alpha 
P^\beta (\epsilon^\gamma (\epsilon'^*\cdot n)+ 
\epsilon'^{*\gamma}(\epsilon\cdot n))(M^2-t) }{4M^2(1+\xi)(P\cdot n)^2}
\right].
\end{eqnarray}
The $ H_i^{(\rho \gamma)}$ and
 $\tilde H_i^{(\rho \gamma)}$ TDAs 
can be identified with the
 corresponding GPDs for the deuteron case, $H_i$ and $\tilde H_i$,
which have been defined in Ref. \cite{BCDP}.

\subsection{Baryonic channels}
When the hadron is a baryon, one needs to consider 3 quark correlators (see \cite{PS3}) and
the formulae become more cumbersome. Denoting by $x_{i}, i =1,2,3$ the momentum fractions 
carried by the three exchanged quarks, we have $\Sigma_{i} x_{i} = 2 \xi$. The discussion of the different
domains requires more care since one or two of the exchanged partons may have to be reinterpreted as 
antiquarks. Lack of space prevents us from describing here these TDAs which may become a central 
concept for understanding hard $\bar p p $ collisions at forward angles.

\section{Phenomenology}
Not much is presently known about the TDAs. 
Sum rules relate some of them to transition form factors as the
$\pi \gamma$ form factor which is experimentally well measured. 
There is no {\it forward} limit as in the case of the GPDs.
There may be positivity constrains which give inequalities 
relating TDAs to parton distributions in mesons or photons.
In some cases implying the $\pi$ meson, 
the chiral limit give some useful information, for instance
 the $\gamma \to  \pi$ TDA is  related to the photon
distribution amplitude 
when the meson becomes soft.

When the pion is not soft one can try to approximate TDAs by extracting
 the pole contribution related to the appropriate exchange in 
$t-$channel.
In particular, the use of the effective interaction lagrangian of $\pi$'s
and $\gamma$ leads in  the case of 
$\gamma \to \pi$ axial TDA $A(x,\xi,t)$ to the expression
\begin{equation}
\label{pole}
A(x,\xi,t)=\frac{2f^2_\pi}{m^2_\pi-t}\,\phi_\pi\left( \frac{x+\xi}{2\xi} \right)
\;\theta(\xi\ge x \ge -\xi)\;,
\end{equation}  
approximating $A$ TDA in the ERBL region. The advantage of such 
approach is that it leads to formulas being to some extent model independent,
the disadvantage is their rather restricted domain of applicability.


Some model estimates are possible \cite{BT}. Without having yet performed 
a complete phenomenological study, 
we are confident that the expected orders of magnitudes of the cross sections 
of processes we discuss are amply sufficient to be measurable in feasible
experimental set ups. Whether the mechanism described here will 
be dominant at accessible values of $Q^2$
cannnot be determined from theory in its present state of the art.

In analogy to the case of form factors, one may be more optimistic on
the validity of the factorized description   at fairly small values of
$Q^2$ in the mesonic case than in the baryonic case. Let us stress
however that the phenomenological difficulties of the hard description
of baryonic form factors at accessible values of $Q^2$ \cite{deJager} 
does not prevent
us from being optimistic on the possible dominance of the hard process
in the new case. The physical reason of this optimism  relies in the
following statement : it may be difficult for a baryon to be in its
valence Fock state, but it should be easier for it to be in a higher
Fock state such as $| q q q  \pi \rangle$.

\section{Conclusion}
The TDAs are new objects with interesting properties. Their physics content seems quite rich
and more developments should helpfully follow. Their $x,\xi, t-$ dependences each map out different features of the 
hadronic structure. For instance the (Fourier transformed)  $t-$ dependence maps out 
\cite{Lisboa} the impact parameter of a short transverse size $\bar q q$ pair in a meson.
On the experimental size, we expect sizable cross sections for both the $\gamma^* \gamma \to A B  $
reactions accessible at existing high luminosity electron positron colliders on the one hand, and the
$\bar p N \to \gamma^* \gamma   $ and $\bar p N \to \gamma^* \pi   $ which will be measured 
at the FAIR project in GSI \cite{GSI}. Backward VCS and backward exclusive meson electroproduction experiments
may (and will)  also be analyzed in this framework.
Finally let us note that generalized version of TDAs, with $\gamma$ and 
hadrons in the same (final or initial) state is a natural framework 
for description of the timelike DVCS processes, e.g. $e^+ e_- \to 
\pi^+\,\pi^-\,\gamma\;$ \cite{br}

\vskip.1in
\noindent
Work of L.Sz. is supported by the Polish Grant 1 P03B 028 28. 
He is a Visiting Fellow of the FNRS (Belgium).

\end{document}